# Electromagnetic analysis of coated conductors with ferromagnetic substrates: Novel insights

**Vladimir Sokolovsky [a,1] and Leonid Prigozhin [b]**

[a] Physics Department, Ben-Gurion University of the Negev, Beer-Sheva 84105 Israel.

[b] Blaustein Institutes for Desert Research, Ben-Gurion University of the Negev, Sde Boqer Campus 84990 Israel.

ABSTRACT

Ferromagnetic substrates can significantly influence the electromagnetic response of a coated conductor to an external magnetic field and transport current. This study analyzes this response theoretically using the thin shell integrodifferential model. First, assuming the substrate is strongly magnetic and the superconductor is in the Meissner state, we present the analytical solution in a convenient explicit form. This helps us to analyze the superconducting current density distributions, highlighting their differences from those in conductors with non-magnetic substrates. Secondly, for a superconducting layer characterized by a nonlinear current-voltage relationship and a substrate with a finite field-independent magnetic permeability, we use an effective spectral numerical method to study the unique features of this hybrid superconductor/ferromagnet system, such as magnetization in a parallel external field and the peculiar nonmonotonic variation of loss observed when alternating transport current and parallel field are applied simultaneously. Dynamic losses for the case of a direct transport current and an alternating parallel field are also investigated. It is shown that tuning the phase and amplitude of the applied parallel field relative to those of the transport current can minimize AC losses.

## 1. Introduction

Coated conductors, such as second-generation high-temperature superconducting tapes, are essential components in modern superconducting technologies, including power transmission systems, high-field magnets, and fault current limiters. These conductors typically feature a multilayered architecture, where the superconducting layer is deposited on a buffered metallic substrate. The ferromagnetic Ni-W substrates are often employed [1-3]. The magnetic nature of the substrate can significantly alter the electromagnetic behavior of the conductor in response to applied magnetic fields and transport currents, and influence the AC loss [2, 4-11]. Such substrates can improve the characteristics of an HTS dynamo flux pump [12] or a magnetic trap for cold atoms on a superconducting chip [13].

---

[1] Corresponding author.
   *E-mail address:* sokolovv@bgu.ac.il (V. Sokolovsky)

Understanding the interaction between nonlinear superconducting behavior and substrate magnetization is crucial for technological applications. The analytical results [2, 6, 7] are limited to idealized cases where the substrate has infinite magnetic permeability, and the superconductor is either in the Meissner state or follows the Bean critical-state model. In the latter case, the critical-state regions are assumed to be two symmetric zones formed near the conductor edges; this assumption is not always true. Numerical simulations using the two-dimensional finite element formulations (see, e.g., [10, 14-17]) are often computationally intensive due to the need to discretize the surrounding space and account for the high aspect ratio of the conductor geometry.

In this work, we employ the thin shell approximation to study a coated conductor with a ferromagnetic substrate. We begin by considering a superconductor in the Meissner state on a strongly magnetic substrate. An explicit analytical expression for the sheet current density in the superconductor is derived, which helps us to highlight the key differences from the non-magnetic substrate case. We then account for a nonlinear current-voltage relationship for the superconductor and a finite magnetic permeability of the substrate. Using a highly efficient spectral numerical method, we investigate magnetization behavior, loss characteristics, and the interplay between external parallel fields and transport currents. Our results show that it is possible to significantly reduce AC and dynamic losses by adjusting the applied field in relation to the transport current. This finding could be important for optimizing the use of coated conductors in superconducting devices.

## 2. Thin shell model of a coated conductor with a ferromagnetic substrate

We consider an infinite in the $z$-axis direction conductor and limit our consideration to the superconducting layer and ferromagnetic substrate; usually, other layers do not significantly affect the electromagnetic response.

The substrate material is assumed to be non-conducting with a constant relative magnetic permeability $\mu_r$ (below, we will use the susceptibility $\chi = \mu_r - 1$). The substrate has the cross-section $\{|x| \leq a, |y| \leq \delta/2\}$, where $2a$ and $\delta$ are its width and thickness, respectively. We assume the applied external field $\boldsymbol{h}^e(t, x, y) = (h_x^e, h_y^e)$ and magnetization of the substrate $\boldsymbol{m}(t, x, y) = (m_x, m_y)$ are parallel to the $xy$-plane.

The asymptotic thin shell model of magnetization [18-20] is derived in the limit $\delta \to 0, \chi \to \infty$ while their product $\chi\delta$ remains constant. This model is justified if $a \gg \delta, \chi \gg 1$ and formulated in terms of "surface magnetization" $\sigma$; in our case $\sigma(t, x) = \int_{-\delta/2}^{\delta/2} m_x(t, x, y) \mathrm{d}y$ and can be treated as a scalar function.

The superconducting layer is assumed to be infinitely thin and placed above the substrate, $\{|x| \leq a, y = \delta/2, |z| < \infty\}$. The sheet current density $j(t, x)$ characterizes the distribution of the current parallel to the $z$-axis in this layer.

In the thin shell model, the coated conductor cross-section shrinks to the interval $\{|x| \leq a, y = 0\}$. The integrodifferential equations describing the electromagnetic response of the bilayer (see [21] for the detailed derivation) can be written as follows:

$$(\chi\delta)^{-1}\sigma(t,x) + \partial_x\left(\frac{1}{2\pi}\int_{-a}^{a}\frac{\sigma(t,x')}{x-x'}dx'\right) - \frac{j}{2} = h_x^e, \quad (1)$$

$$\partial_x e = \mu_0 \partial_t\left(h_y^e + \frac{1}{2\pi}\int_{-a}^{a}\frac{j(t,x')}{x-x'}dx' - \frac{\partial_x\sigma}{2}\right), \quad (2)$$

$$\int_{-a}^{a} j(t,x)dx = I(t), \qquad \sigma(t,\pm a) = 0. \quad (3)$$

Here $e(x,t)$ is the electric field in the superconducting layer, $I(t)$ is the applied transport current, $\mu_0$ - the magnetic permeability of vacuum, and the singular integrals should be interpreted in the sense of the Cauchy principal value.

The term $-\partial_x\sigma/2$ in the equation (2) describes the substrate-magnetization-induced normal field acting upon the superconducting layer. This non-uniform field modifies the electromagnetic response of the superconductor to an applied field or transport current and, in particular, is the reason for superconducting layer magnetization in a parallel external field [13, 21].

The main interest for applications presents a superconductor in the mixed state, characterized by a nonlinear current-voltage relation $e = e(j)$. In this work, we assume the power law,

$$e = e_0 \left|\frac{j}{j_c}\right|^{n-1} \frac{j}{j_c}, \quad (4)$$

where $e_0 = 10^{-4}$ Vm$^{-1}$, the power $n$ and the critical sheet current density $j_c$ are, for simplicity, assumed constants. The problem (1)-(4) is evolutionary, and, in our examples, we will assume $j(0,x) = 0$. The accumulating loss in the superconducting layer per unit of its length, $Q$, evolves as

$$\frac{dQ}{dt} = \int_{-a}^{a} ej \, dx. \quad (5)$$

The magnetic field outside the conductor can be expressed in terms of $j$ and $\sigma$ [21]. As in that work, we will use the dimensionless variables,

$$\tilde{j} = \frac{j}{j_c}, \quad \tilde{\sigma} = \frac{\sigma}{aj_c}, \quad \tilde{\mathbf{h}} = \frac{\mathbf{h}}{j_c}, \quad \tilde{e} = \frac{e}{e_0},$$

$$(\tilde{x}, \tilde{y}) = \frac{(x,y)}{a}, \quad \tilde{t} = \frac{t}{t_0}, \quad \tilde{I} = \frac{I}{I_c}, \quad \tilde{Q} = \frac{Q}{\mu_0 a^2 j_c^2}, \quad (6)$$

where $t_0 = a\mu_0 j_c / e_0$ and the critical current $I_c = 2aj_c$. In the dimensionless form (with the sign "~" omitted), equations (1)-(5) become

$$\kappa^{-1}\sigma(t,x)+\partial_x\left(\frac{1}{2\pi}\int_{-1}^{1}\frac{\sigma(t,x')}{x-x'}dx'\right)-\frac{j(t,x)}{2}=h_x^e(t,x), \tag{7}$$

$$\partial_x e(t,x)=\partial_t\left(h_y^e+\frac{1}{2\pi}\int_{-1}^{1}\frac{j(t,x')}{x-x'}dx'-\frac{\partial_x\sigma(t,x)}{2}\right), \tag{8}$$

$$\int_{-1}^{1}j(t,x)dx=2I(t),\quad \sigma(t,\pm 1)=0, \tag{9}$$

$$e=|j|^{n-1}j, \tag{10}$$

$$\frac{dQ}{dt}=\int_{-1}^{1}ej\,dx. \tag{11}$$

Here $\kappa=\chi\delta/a$ is the dimensionless parameter fully characterizing the magnetic substrate in our model. The magnetic susceptibility $\chi$ is zero for nonmagnetic materials. For ferromagnetic substrates, it can vary from tens to thousands, and even more [1]. For instance, for a coated conductor width of 10 mm and a substrate thickness of 100 μm we have $\delta/a=0.02$. So, the parameter $\kappa$ may vary over a wide range. As $\kappa\to 0$, the solution tends to the known analytical solution for coated conductors with a nonmagnetic substrate [22]. Our previous simulations [21] showed that for, say, all $\kappa>10$ the solutions are practically the same. It seems beneficial to first study a simpler version of this model for which the problem can be solved analytically.

### 3. Superconductor in the Meissner state on a strongly magnetic substrate

If the superconductor is in the Meissner state, the current-voltage relation (4) should be replaced by the condition that the magnetic field normal to the superconducting layer is zero. The model lacks a characteristic critical sheet current density $j_c$ and, to use the dimensionless variables (6), we define this value arbitrarily, e.g., by setting $j_c=1\,\text{A/m}$. To simplify the model further, let us assume the substrate susceptibility is very high, $\kappa\gg 1$, and neglect the first term in (7). Using integration by parts, we can also rewrite the second term of this equation,

$$\partial_x\left(\frac{1}{2\pi}\int_{-1}^{1}\frac{\sigma(t,x')}{x-x'}dx'\right)=\frac{1}{2\pi}\int_{-1}^{1}\frac{\partial_{x'}\sigma(t,x')}{x-x'}dx'.$$

The resulting model is quasi-stationary and, in dimensionless form, can be written as

$$\frac{1}{2\pi}\int_{-1}^{1}\frac{\partial_{x'}\sigma(t,x')}{x-x'}dx'-\frac{j(t,x)}{2}=h_x^e, \tag{12}$$

$$\frac{1}{2\pi}\int_{-a}^{a}\frac{j(t,x')}{x-x'}dx'-\frac{\partial_x\sigma}{2}=-h_y^e, \tag{13}$$

$$\int_{-1}^{1}j(t,x)dx=2I(t),\quad \sigma(t,\pm 1)=0. \tag{14}$$

An analytical solution has been presented in [21] for the transport current problem; here, we extend this solution to the presence of a uniform external field (Appendix):

$$j = I\frac{(g^+ + g^-)}{\sqrt{2\pi}} + h_x^e \frac{1}{\sqrt{2}}\left[x(g^- - g^+) - \frac{g^+ + g^-}{2}\right] + h_y^e \frac{1}{\sqrt{2}}\left[x(g^+ + g^-) + \frac{g^+ - g^-}{2}\right], \quad (15)$$

$$\partial_x \sigma = I\frac{g^+ - g^-}{\sqrt{2\pi}} + h_x^e \frac{1}{\sqrt{2}}\left[-x(g^+ + g^-) + \frac{g^- - g^+}{2}\right] + h_y^e \frac{1}{\sqrt{2}}\left[x(g^+ - g^-) + \frac{g^+ - g^-}{2}\right], \quad (16)$$

where $g^+ = (1+x)^{-3/4}(1-x)^{-1/4}$, $g^- = (1+x)^{-1/4}(1-x)^{-3/4}$.

We note that Mawatari [6] proposed another approach to solving the problem in the infinite substrate permeability limit ($\mu_r \to \infty$). Using functions of complex argument, he found the magnetic field outside the conductor. The sheet current density $j$ can be sought as the jump of the tangential component of this field on the conductor. Our explicit solution (15)-(16) is more convenient.

The sheet current densities and the substrate magnetization-induced normal fields $h^m = -\partial_x \sigma / 2$ for the three characteristic cases are shown in Fig. 1. For an arbitrary external field and transport current, the solution is a linear combination of these three.

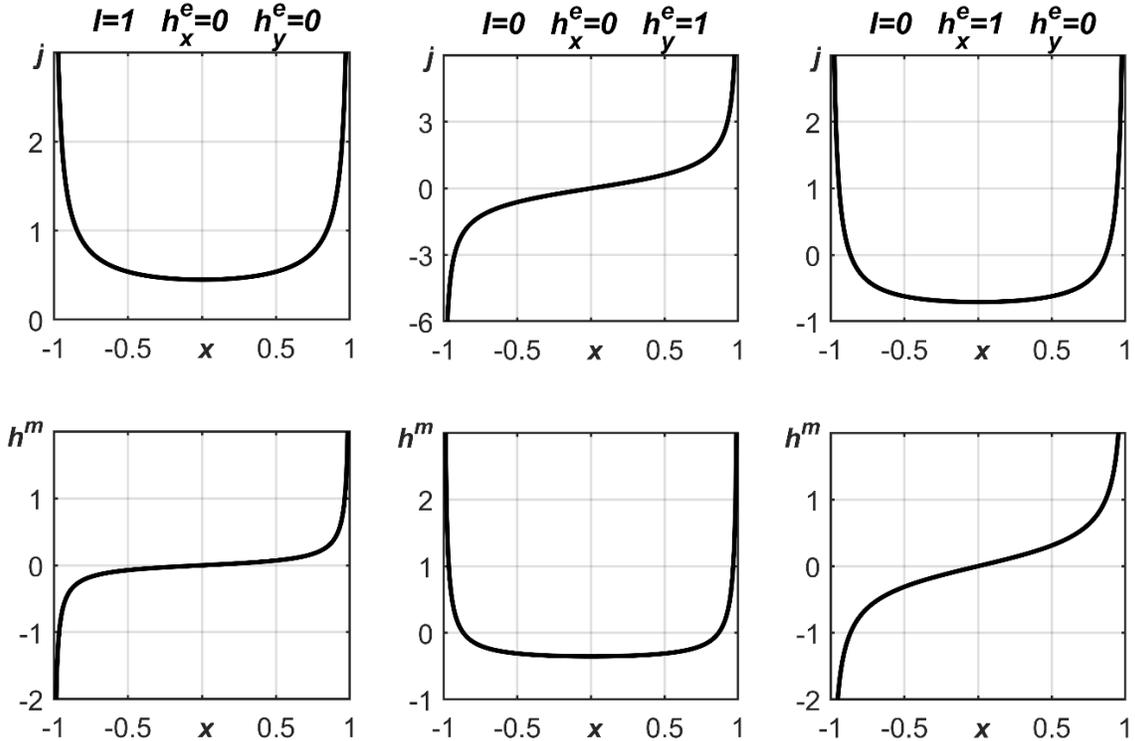

**Fig. 1.** Analytical solutions for the Meissner state and $\kappa \to \infty$. Top: sheet current density distributions, bottom: substrate magnetization-induced non-uniform normal fields $h^m = -\partial_x \sigma / 2$.

Generally, the sheet current density tends to plus or minus infinity at the strip edges as $(1-|x|)^{-3/4}$ and this is faster than $(1-|x|)^{-1/2}$ for a strip in the Meissner state on a non-magnetic substrate [22]. The sign of $j$ near $x=1$ is the sign of $I/\pi + (h_x^e + h_y^e)/2$, and near $x=-1$ is that of $I/\pi + (h_x^e - h_y^e)/2$, provided that these expressions are nonzero. If either of the two expressions vanishes, then $j \sim (1-|x|)^{-1/4}$ near the corresponding edge.

The main interest is the case of a nonzero parallel field $h_x^e$: otherwise, the distributions of $j$ are qualitatively similar to those studied for non-magnetic substrates. If $h_x^e > 0, h_y^e = 0, I = 0$, the sheet current density tends to $+\infty$ at the strip edges and is negative near the strip center, where the minimum, $j(0) = -h_x^e/\sqrt{2}$, is attained. If also $I \neq 0$ the distribution remains even, the extremum at $x=0$ becomes $j(0) = (2I/\pi - h_x^e)/\sqrt{2}$; it can also be a local maximum (Fig. 2, left). For $h_y^e = 1, I = 0$, and several values of $h_x^e$ the sheet current density distributions are shown in Fig. 2, right.

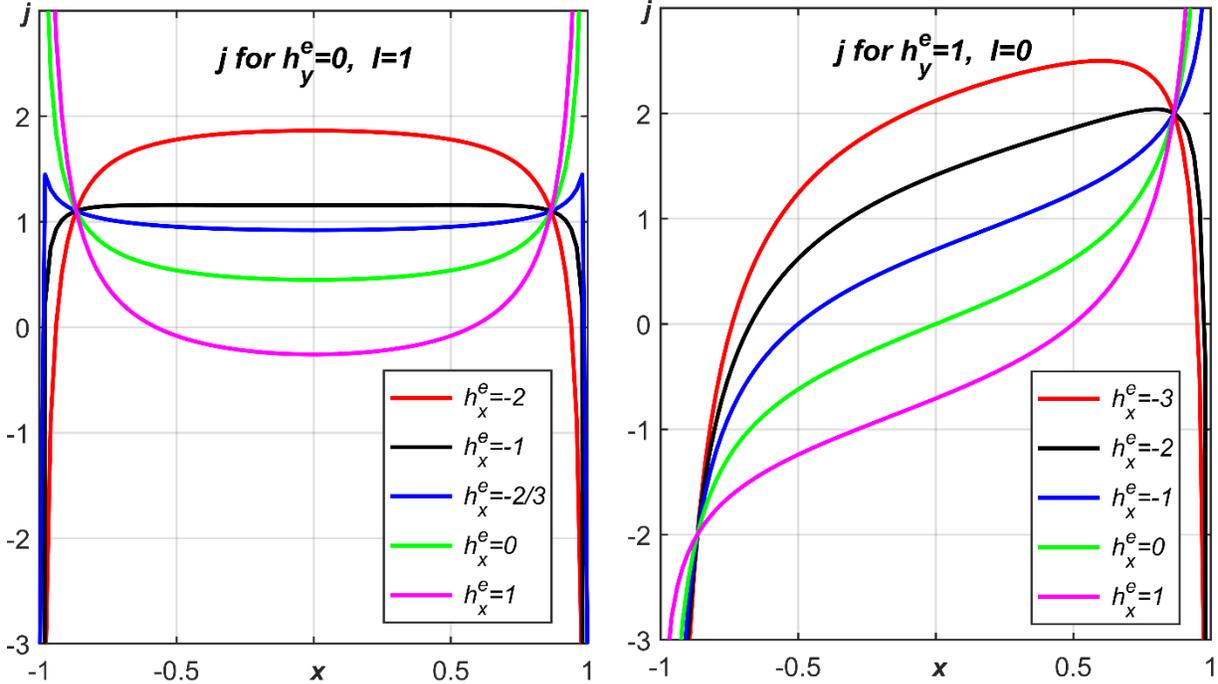

**Fig. 2.** Sheet current density distributions for different combinations of the applied field and transport current; analytical solutions for the Meissner state and $\kappa \to \infty$.

The obtained analytical solutions help to understand the response of a non-idealized conductor to weak applied fields and currents. These solutions can also help to understand the electromagnetic behavior of a coated conductor if the superconducting layer is characterized by a nonlinear current-voltage relation, like (4) or its $n \to \infty$ limit, the Bean critical-state model, and the magnetic permeability of a ferromagnetic substrate is finite. Unlike the sheet current density in the model (12)-(14) considered above, solutions of the equations (7)-(10) are bounded. It can be expected that the sheet current density is close to the critical values, $\pm 1$, where the corresponding solutions to (12)-(14) are large, e.g., near the strip edges. In these areas of the superconductor, the main loss occurs. If the applied parallel field $h_x^e$ is strong, such a region can appear near the central axis of the strip too. By applying simultaneously, e.g., a negative parallel field $h_x^e$ and a positive transport current $I$ it is possible to obtain a close-to-uniform subcritical sheet current density distribution in the central part of the strip and only small critical sheet current density regions near the strip edges (see the example and Fig. 5 below).

## 4. Superconductor with a nonlinear current-voltage relation

For a coated conductor with a transport current or in a uniform external field, functions of complex arguments were used to find the magnetic field outside the conductor in [6] and [16]. These analytical solutions were obtained assuming the superconductor obeys the Bean critical-state model, the magnetic permeability of the substrate is infinite, and there are precisely two symmetric critical regions propagating from the superconducting strip edges. Despite the importance of these exact solutions, the first two assumptions limit their applicability, and if, e.g., a third critical-state region appears or the external field and transport current act simultaneously, no analytical solution is available.

Numerical simulations can be employed to model the evolving electromagnetic response of the conductor to an arbitrary combination of the varying with time applied magnetic field and transport current. The two-dimensional finite element methods were used for this in a number of works (see, e.g., [2, 9, 16, 23]). The necessity to consider the surrounding space and the very high aspect ratio of the thin layers seem to be the disadvantages of this approach.

Here, we solve the integrodifferential system (7)-(10) numerically using the highly accurate spectral method developed in [21]. The method employs spatial approximation of unknowns by Chebyshev polynomial expansions; this allows for exact analytical evaluation of the singular integrals in the equations (7) and (8). The method of lines and a standard ordinary differential equation solver are used to integrate the system in time. Due to the thin shell approximation, the model is one-dimensional, and our method is efficient: solving the examples presented below using a mesh of 200 nodes on a PC takes 2–15 seconds.

In our first example (Fig. 3), we chose the high power, $n = 100$, in the power law (10) to mimic the rate-independent Bean critical-state model. We also set $\kappa = 15$ to model a strongly magnetic substrate. We assumed $I = 0, h_y^e = 0$. The parallel magnetic field first grows from zero up to $h_x^e = 6$ (Fig. 3, top), then decreases and becomes negative (Fig. 3, bottom).

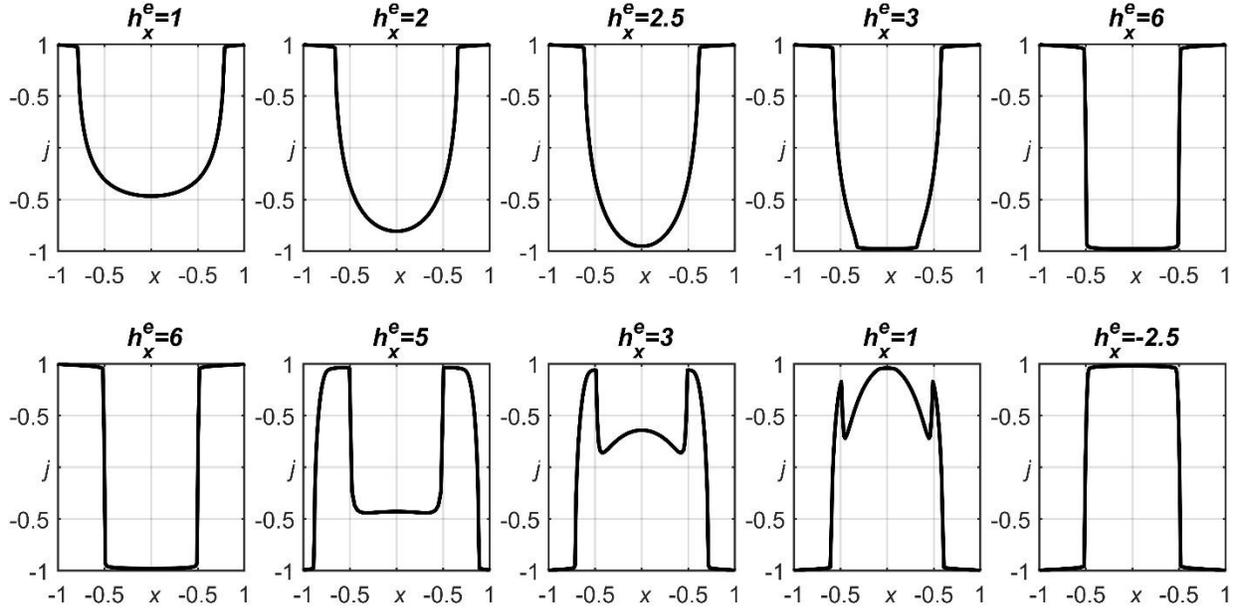

**Fig. 3.** Sheet current density in the Bean-like model of the superconducting layer ($n=100$) on a strongly magnetic substrate ($\kappa=15$). The parallel external field $h_x^e$ first grows (top), then decreases and becomes negative (bottom).

As the field $h_x^e$ starts to grow, the critical state regions with $j \approx 1$ appear near the strip edges and propagate into the strip. The magnitude of the negative sheet current density in the strip middle also increases, and, as the external field reaches $h_x^e \approx 2.6$, there appears and grows an additional critical state region with $j \approx -1$. For $h_x^e > 5$ almost the whole strip is in the critical state $|j| \approx 1$. When the external field decreases, the critical regions with $j \approx -1$ begin to propagate into the strip from its edges, while in the central region $j$ becomes subcritical, then changes the sign; the critical state in the whole strip is, for a strongly magnetic substrate, almost reached already at $h_x^e = -2.5$.

In our following example (Fig. 4), we simulated the coated conductor response to a combination of the transport current $I = 0.25\min(t,1)$ and a parallel external field, either $h_x^e = \max(t-1,0)$ or $h_x^e = -\max(t-1,0)$. Here, we compare the solutions for strongly ($\kappa=15$) and moderately ($\kappa=1$) ferromagnetic substrates and also for the Bean-like current-voltage relation and the power relation with the low power value $n=15$ for the superconductor. The number of the critical-state regions and the sign of $j$ within them depend on the magnitude and direction of the parallel external field.

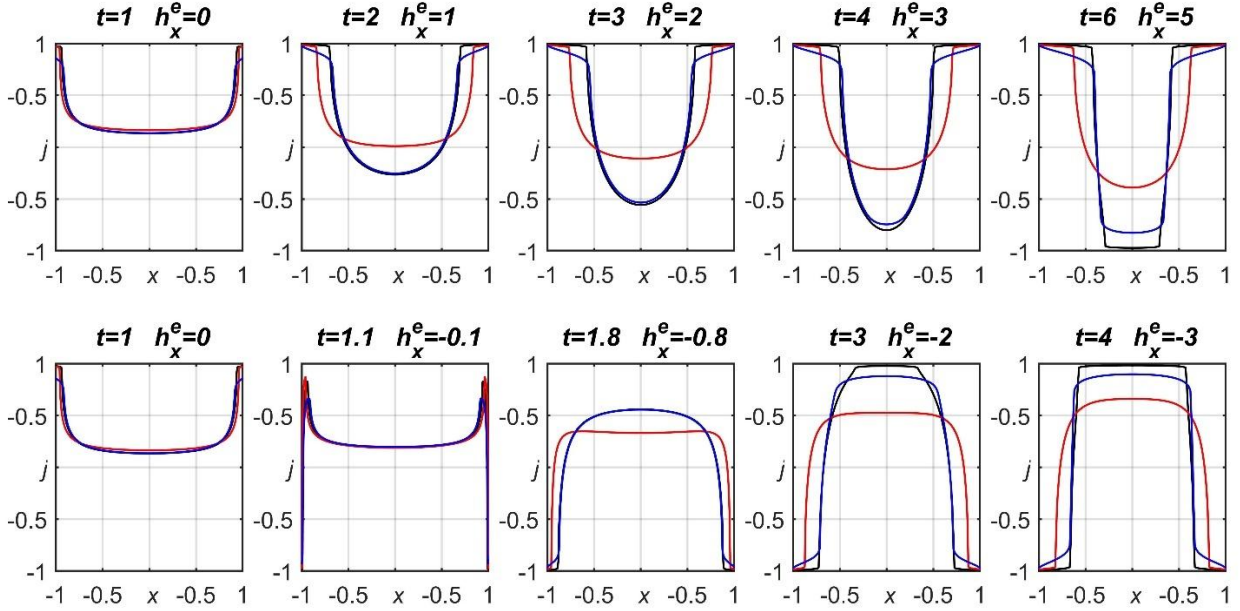

**Fig. 4.** Sheet current density distributions for the transport current $I = 0.25\min(t,1)$ and a parallel magnetic field. Top: $h_x^e = \max(t-1,0)$; bottom: $h_x^e = -\max(t-1,0)$. Model parameters: $\kappa = 15, n = 100$ (black lines), $\kappa = 1, n = 100$ (red lines), and $\kappa = 15, n = 15$ (blue lines).

By adjusting the parallel external field $h_x^e$ it is possible to run even a strong transport current with the sheet current density being subcritical, $|j|<1$, almost everywhere (fig. 5). This suggests that AC losses can be decreased if the ratio of a parallel magnetic field to transport current is negative. As is shown below (Fig. 7), this effect is observed in a range of the ratio values depending on $\kappa$.

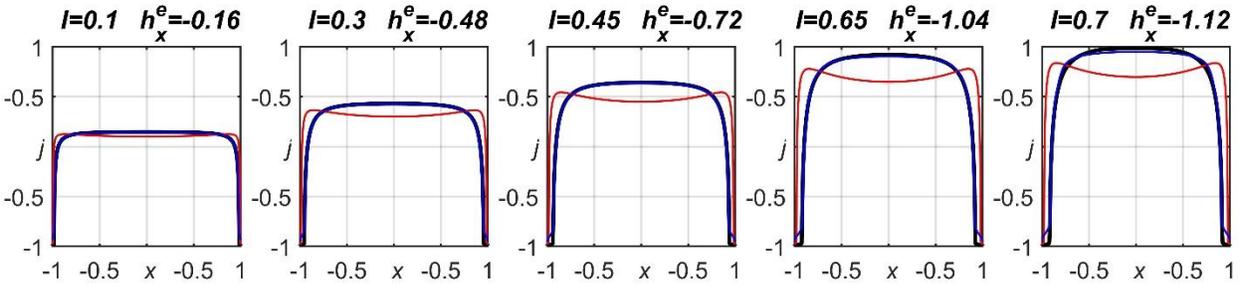

**Fig. 5.** Adjusting the parallel external field to keep the sheet current density subcritical. In this example, $I = t, h_x^e = -1.6t$. The black lines ($\kappa = 15, n = 100$) almost coincide with the blue lines ($\kappa = 15, n = 15$); red lines correspond to $\kappa = 1, n = 100$.

Our results (Figs. 4-5) suggest that at the chosen time scale $t_0 = a\mu_0 j_c / e_0$, the power $n$ in (4) has only a weak influence on the current density. For a typical coated conductor ($2a = 1$ cm, $J_c = 3\cdot10^4$ A/m) $t_0$ is 1.9 s. The nonlinear diffusion of magnetic flux, determined by (4), becomes more significant for slower variations of the external field and/or transport current. To demonstrate this, we set $h_y^e = t/10$, $h_x^e = k_1 h_y^e$, $I = k_2 h_y^e$ with different combinations of the constants $k_1$ and $k_2$ (Fig. 6). Now the influence of the power $n$ is more pronounced. The asymmetric sheet current density distributions, even for large $n$ and $\kappa$ not represented by the known analytical solutions, appear when both external field components or both the transport current and the normal field component are non-zero.

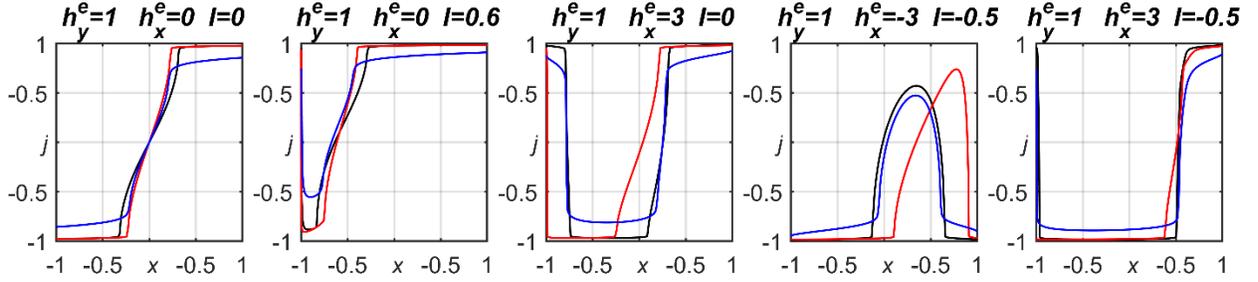

**Fig. 6.** Sheet current density for different combinations of the slowly changing external field and transport current. Model parameters: $\kappa = 15$, $n = 100$ (black lines), $\kappa = 1$, $n = 100$ (red lines), and $\kappa = 15$, $n = 15$ (blue lines).

Dependences of AC losses in coated conductors with a ferromagnetic substrate on the transport current or the external field have been studied, e.g., in [2, 4, 5, 8, 10, 11, 16, 21]. Less attention was paid to the influence on AC losses of the current and field applied simultaneously. As was noted above, not *a priori* apparent effect can be produced by a combination of the AC parallel field and transport current due to specific properties of the superconducting current density distributions (Fig. 5).

Let us set $n = 30$, $I = I_0 \sin(2\pi t / T)$, $h_x^e = h_0 \sin(2\pi t / T + \varphi)$, $h_y^e = 0$ with $\varphi = 0$ or $\varphi = \pi$ and $T = 0.01$. For a typical $t_0 \approx 2$ s, the latter value corresponds to the frequency 50 Hz. Using (11), we computed the AC loss per unit of length, $Q$, during the second period, $T \le t \le 2T$, for different amplitudes $h_0$, $I_0$, and two values of $\kappa$: 15 for a strongly magnetic substrate and 0.5 for a moderately magnetic one. For the in-phase variation of $I$ and $h_x^e$ ($\varphi = 0$), an expected behavior of the AC loss is observed: the loss grows monotonically with both amplitudes, $I_0$ and $h_0$, and also increases with the value of $\kappa$ (Fig. 7, left).

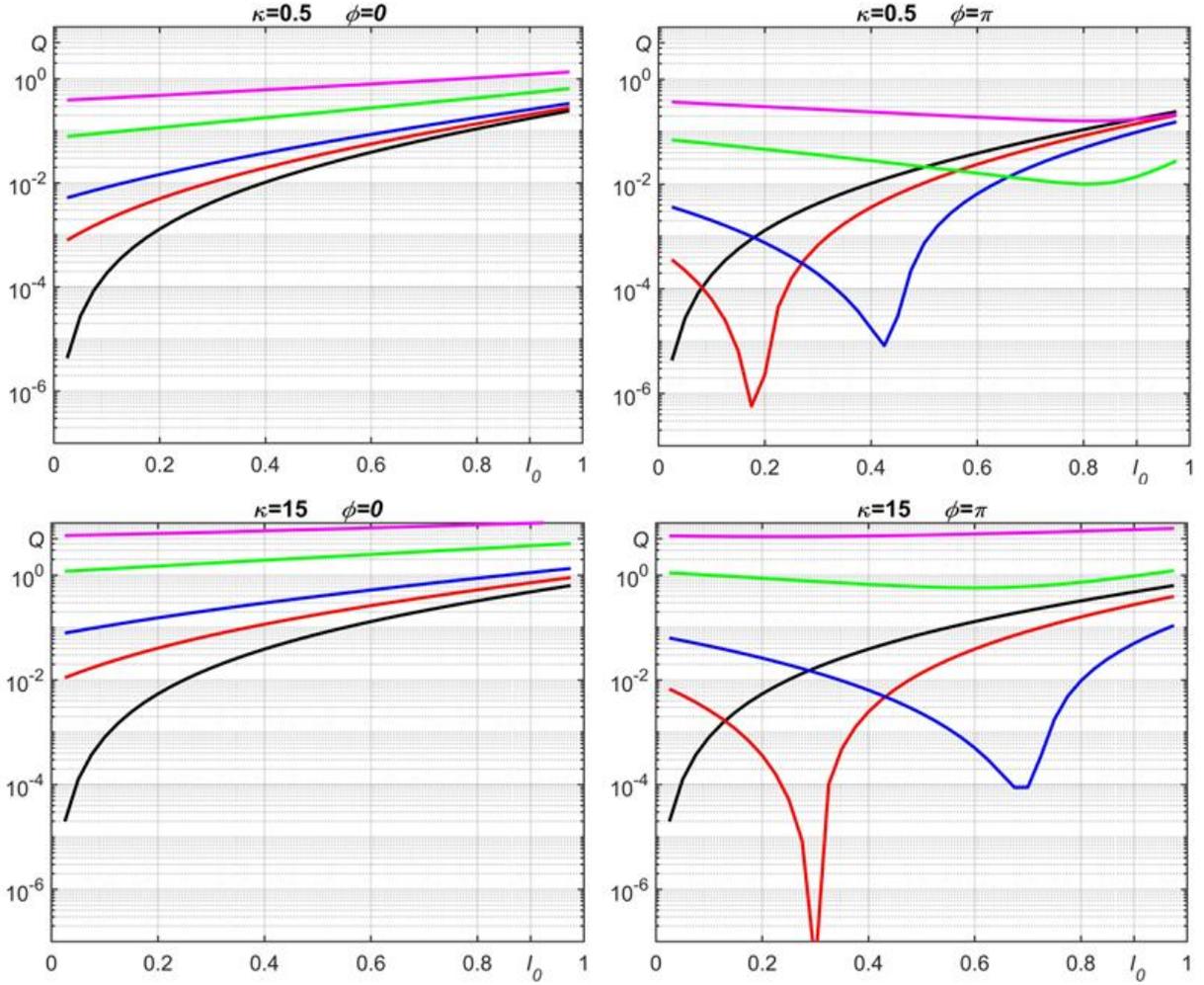

**Fig. 7.** AC loss per unit of length per period for simultaneously changing $I = I_0 \sin(2\pi t/T)$ and $h_x^e = h_0 \sin(2\pi t/T + \varphi)$ with $T = 0.01$, $\varphi = 0$ (left) or $\pi$ (right) and two values of $\kappa$: 0.5 (top) and 15 (bottom). Curves: $h_0 = 0$ - black, $h_0 = 0.2$ - red, $h_0 = 0.5$ - blue, $h_0 = 2$ - green, $h_0 = 5$ - magenta.

However, if $\varphi = \pi$ the loss can be several magnitudes lower, provided the field amplitude is properly adjusted to the current amplitude (Fig. 7, right). In such a case, as is shown in Fig. 5, the sheet current density remains subcritical everywhere except the two tiny regions at the strip edges, where the loss occurs. The effect should be observed for realistic values of the applied fields and currents (taking as a typical characteristic value $j_c \sim 3\cdot 10^4$ A/m, we find that $h_x^e = 1$ corresponds to about 0.04 T).

Coated conductors carrying a direct current are often subjected to the action of an alternating magnetic field, which leads to energy dissipation, called dynamic loss. The influence of a magnetic substrate on the dynamic loss has been investigated for the fields normal to the conductor strip (see [4] and the references therein). To examine the effect of a parallel alternating field, we assumed $n = 30$, $I = 0.75$, $h_y^e = 0$, and a DC-biased $h_x^e = A + B\sin(2\pi t/T)$ with $T = 0.01$. In this simulation, we computed the dynamic losses per period per unit length (Fig. 8) for different values of $A, B$ and $\kappa$, established after the transient time. In agreement with our previous results (Fig. 7), since $I > 0$, the dynamic losses for $A < 0$ are lower than those for $A = 0$, which are less than those for $A > 0$.

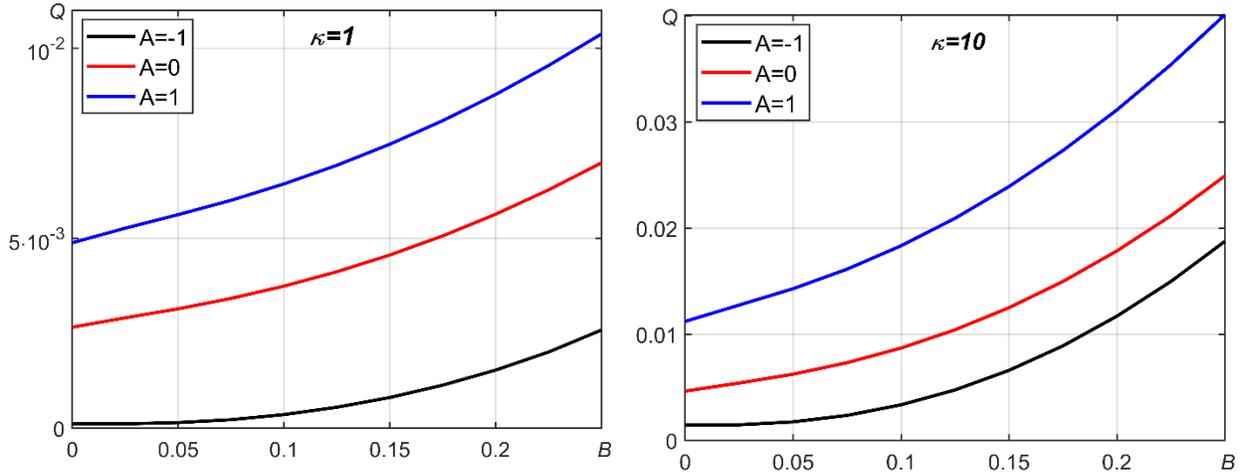

**Fig. 8.** Dynamic loss per period per unit length for $I = 0.75$ and DC-biased alternating parallel external field $h_x^e = A + B\sin(2\pi t/T)$. Left: $\kappa = 1$; right: $\kappa = 10$.

## 5. Conclusion

We have investigated the electromagnetic behavior of coated conductors with ferromagnetic substrates using a combination of analytical and numerical techniques within the thin shell approximation. An explicit analytical solution was obtained for a superconductor in the Meissner state placed on a highly magnetic substrate, revealing distinctive features of the sheet current density distribution compared to the non-magnetic case. For a more realistic case involving a nonlinear current-voltage relation and a substrate with a finite magnetic permeability, we employed an accurate and efficient spectral numerical method to simulate the response of the coated conductor to various combinations of the applied fields and transport currents.

A ferromagnetic substrate significantly modifies current density distributions, particularly in the presence of a parallel external field. Importantly, we showed that AC and dynamic losses can be reduced by tuning the amplitude and phase of the applied field relative to the transport current in order to maintain subcritical current densities across most of the superconductor.

These insights provide a potential pathway for improving the efficiency of superconducting devices based on coated conductors.

Although in our simulations, the magnetic permeability of the substrate was assumed constant, the employed thin shell model can be extended to the case of a field-dependent permeability, provided the condition $\mu_r \gg 1$ remains valid. Furthermore, as was shown in [21], for any value of $\kappa = (\mu_r - 1)\delta/a > 10$ the electromagnetic response of a coated conductor is practically the same. Hence, while this condition holds, the variation of $\mu_r$ can be ignored.

## Appendix: Analytical solution of the system (12)-(14)

Denoting $u^+ = j + \partial_x \sigma$, $u^- = j - \partial_x \sigma$ and using (12)-(14), we obtain an integral equation of the second kind supplemented by the integral transport current constraint for each of these variables:

$$u^{\pm} \mp \frac{1}{\pi} \int_{-1}^{1} \frac{u^{\pm}(t,x')}{x-x'} dx' = f^{\pm}, \quad \int_{-1}^{1} u^{\pm}(t,x) dx = 2I, \tag{17}$$

where $f^+ = 2(h_y^e - h_x^e)$ and $f^- = -2(h_x^e + h_y^e)$. For a superconducting strip in the Meissner state, the sheet current density $j$ is infinite at the strip edges; hence, this is true also for $u^+$ and $u^-$. Solutions to (17), unbounded at the interval ends, are [24]:

$$u^{\pm} = \frac{f^{\pm}(t,x)}{2} \pm \frac{g^{\pm}(x)}{2\pi} \int_{-1}^{1} \frac{f^{\pm}(t,x')}{g^{\pm}(x')(x-x')} dx' + C g^{\pm}(x) \tag{18}$$

with $g^+ = (1+x)^{-3/4}(1-x)^{-1/4}$, $g^- = (1+x)^{-1/4}(1-x)^{-3/4}$, and the constants $C$ determined by the integral (transport current) constraints. If the applied field is uniform, we have

$$u^{\pm} = \frac{f^{\pm}(t)}{2} \pm \frac{g^{\pm}(x) f^{\pm}(t)}{2\pi} \int_{-1}^{1} \frac{dx'}{g^{\pm}(x')(x-x')} + C g^{\pm}(x). \tag{19}$$

We compute the integral in (19) using the formula (see [25], 3.228.3)

$$I^{\alpha} = \int_{-1}^{1} \frac{(1+x')^{1-\alpha}(1-x')^{\alpha}}{x-x'} dx' = -\pi \cot(\alpha\pi)(1+x)^{1-\alpha}(1-x)^{\alpha} + 2B(\alpha, 2-\alpha) \cdot {}_2F_1\left(-1, 1; 1-\alpha; \frac{1-x}{2}\right)$$

with $\alpha = 1/4$ for $u^+$ and $\alpha = 3/4$ for $u^-$. Here, the hypergeometric function ([25], 9.100)

$${}_2F_1(a,b;c;z) = 1 + \frac{ab}{c} \frac{z}{1!} + \frac{a(a+1)b(b+1)}{c(c+1)} \frac{z^2}{1!} + \ldots, \text{ so } {}_2F_1\left(-1, 1; 1-\alpha; \frac{1-x}{2}\right) = 1 - \frac{1-x}{2(1-\alpha)}. \text{ For the}$$

Beta function, we have $B(1/4, 7/4) = 3\pi\sqrt{2}/4$ and $B(3/4, 5/4) = \pi\sqrt{2}/4$. Hence,

$$I^{1/4} = -\frac{\pi}{g^+} + \frac{\pi}{\sqrt{2}}(2x+1), \quad I^{3/4} = \frac{\pi}{g^-} + \frac{\pi}{\sqrt{2}}(2x-1).$$

Substituting these into (19) we obtain $u^+ = f^+ x g^+ / \sqrt{2} + C^+ g^+$, $u^- = -f^- x g^- / \sqrt{2} + C^- g^-$

and use the transport current constraint in (17) to find

$$C^\pm = \left[2I \mp \left(f^\pm/\sqrt{2}\right)\int_{-1}^{1} xg^\pm(x)\mathrm{d}x\right]/\int_{-1}^{1} g^\pm(x)\mathrm{d}x.$$

Here (see [25], 3.196.3) $\int_{-1}^{1} g^+(x)\mathrm{d}x = \int_{-1}^{1} g^-(x)\mathrm{d}x = B(3/4, 1/4) = \pi\sqrt{2}$. We also found that $\int_{-1}^{1} xg^+(x)\mathrm{d}x = -\int_{-1}^{1} xg^-(x)\mathrm{d}x = -\pi/\sqrt{2}$. Thus, $C^\pm = \left(2I/\pi + f^\pm/2\right)/\sqrt{2}$. Finally,

$$j = (u^+ + u^-)/2, \quad \partial_x \sigma = (u^+ - u^-)/2.$$

**CRediT authorship contribution statement**

**V. Sokolovsky:** Writing – original draft & editing, Conceptualization, Methodology, Investigation, Software. **L. Prigozhin:** Writing – review & editing, Investigation, Software.

**Funding statement**

This research received no specific grant from funding agencies in the public, commercial, or not-for-profit sectors.

**Declaration of competing interest**

The authors declare that they have no known competing financial interests or personal relationships that could have appeared to influence the work reported in this paper.

**Data availability**

All data supporting this study's findings are included within the article.